\documentclass[journal=jacsat,manuscript=article]{achemso}

\usepackage[version=3]{mhchem} 
\usepackage[dvipsnames]{xcolor}
\usepackage{ulem}
\usepackage{float}
\usepackage{amsmath}

\newcommand{\comment}[1]{{\color{black}#1}}

\author{Qi You\textsuperscript{\#}}  
\affiliation[Chemistry]
{State Key Laboratory of Physical Chemistry of Solid Surface, iChEM, College of Chemistry and Chemical Engineering, Xiamen University, Xiamen, 361005, China}

\author{Yan Sun\textsuperscript{\#}}
\affiliation[Chemistry]
{State Key Laboratory of Physical Chemistry of Solid Surface, iChEM, College of Chemistry and Chemical Engineering, Xiamen University, Xiamen, 361005, China}
\alsoaffiliation[dicp]
{State Key Laboratory of Catalysis, iChEM, Dalian Institute of Chemical Physics, Chinese Academy of Sciences, Dalian, 116023, China}

\author{Feng Wang}
\affiliation[IKKEM]
{Laboratory of AI for Electrochemistry (AI4EC), Tan Kah Kee Innovation Laboratory (IKKEM), Xiamen, 361005, China}

\author{Jun Cheng}
\email{chengjun@xmu.edu.cn}
\affiliation[Chemistry]
{State Key Laboratory of Physical Chemistry of Solid Surface, iChEM, College of Chemistry and Chemical Engineering, Xiamen University, Xiamen, 361005, China}
\alsoaffiliation[IKKEM]
{Laboratory of AI for Electrochemistry (AI4EC), Tan Kah Kee Innovation Laboratory (IKKEM), Xiamen, 361005, China}
\alsoaffiliation[AI]
{Institute of Artificial Intelligence,  Xiamen University, Xiamen, 361005, China}

\author{Fujie Tang}
\email{tangfujie@xmu.edu.cn}
\affiliation[Pen-Tung]
{Pen-Tung Sah Institute of Micro-Nano Science and Technology, Xiamen University, Xiamen, 361005, China}
\alsoaffiliation[IKKEM]
{Laboratory of AI for Electrochemistry (AI4EC), Tan Kah Kee Innovation Laboratory (IKKEM), Xiamen, 361005, China}
\alsoaffiliation[AI]
{Institute of Artificial Intelligence,  Xiamen University, Xiamen, 361005, China}

\title{Decoding the Competing Effects of Dynamic Solvation Structures on Nuclear Magnetic Resonance Chemical Shifts of Battery Electrolytes via Machine Learning}

\abbreviations{First-principles, NMR, Machine Learning}

\begin{document}

\begin{tocentry}

\begin{center}
\includegraphics{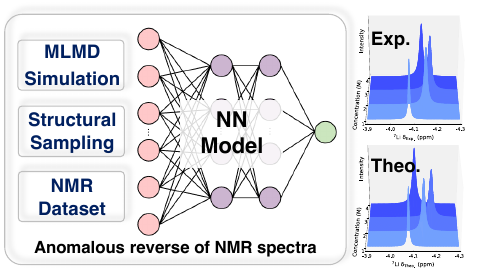}
\end{center}

\end{tocentry}

\begin{abstract}
Understanding the solvation structure of electrolytes is critical for optimizing the electrochemical performance of rechargeable batteries, as it directly influences properties such as ionic conductivity, viscosity, and electrochemical stability. The highly complex structures and strong interactions in high-concentration electrolytes make accurate modeling and interpretation of their ``structure-property" relationships even more challenging with spectroscopic methods. In this study, we present a machine learning-based approach to predict dynamic $^7$Li NMR chemical shifts in LiFSI/DME electrolyte solutions. Additionally, we provide a comprehensive structural analysis to interpret the observed chemical shift behavior in experiments, particularly the abrupt changes in $^7$Li chemical shifts at high concentrations. Using advanced modeling techniques, we quantitatively establish the relationship between molecular structure and NMR spectrum, offering critical insights into solvation structure assignments. Our findings reveal the coexistence of two competing local solvation structures that shift in dominance as electrolyte concentration approaches the concentrated limit, leading to an anomalous reverse of $^7$Li NMR chemical shift in the experiment. This work provides a detailed molecular-level understanding of the intricate solvation structures probed by NMR spectroscopy, leading the way for enhanced electrolyte design.
  
\end{abstract}

\section{Introduction}
Electrolytes facilitate the transfer of ions between the anode and cathode, and their solvation structure can influence various electrochemical properties, such as conductivity,\cite{yoshida_OxidativeStability_2011, chou_Asymmetric_2023} viscosity,\cite{xiao_Insights_2023} and battery performance, including the electrochemical stable potential window (ESPW),\cite{Peljo_Eletrochemical_2018, xiao_Insights_2023} coulombic efficiency (CE)\cite{ko_Electrode_2022}, and cycling reversibility.\cite{cheng_Emerging_2022, sun_Sorbitol_2024} The relative strength of interactions between the cation–solvent and cation-anion can give rise to different types of solvation structures in the electrolytes.\cite{yamada_Superconcentrated_2015, jiang_Inhibiting_2021, ravikumar_Effect_2018, callsen_Solvation_2017, zhang_Constructing_2024, efaw_Localized_2023, yao_Applying_2022} These structures significantly affect the formation of electrode-electrolyte interface,\cite{cheng_Emerging_2022, gu_Insituraman_2023, liu_DeterminationPZC_2024, chen_Atomic_2020, zhang_Functional_2023} the desolvation and diffusion process,\cite{Gupta_Design_2020, hou_Fundamentals_2020} and reaction mechanism.\cite{xiao_Insights_2023, zhang_Radical_2018, jia_Review_2023} In recent years, highly concentrated electrolytes (HCEs) and localized highly concentrated electrolytes (LHCEs) are widely employed in electrochemical energy storage systems\cite{jiang_Inhibiting_2021, cao_Localized_2021}, where the intricate solvation structures and strong interactions between solutes and solvents play a pivotal role in achieving high energy storage densities.\cite{yamada_Superconcentrated_2015, Lukatskaya_Concentrated_2018, Chen_63m_2020} These factors make the accurate modeling and interpretation of their ``structure-property" relationships particularly challenging. Consequently, revealing the solvation structures of electrolytes is the central topic in understanding these processes of interest.\cite{ming_Molecular_2019, gupta_Influence_2020, chen_HighEfficiency_2018}

In the experiments, there have been several characterization methods used to illustrate solvation structures, including vibrational and nuclear magnetic resonance (NMR) spectroscopies. Specifically, Raman spectroscopy has often been employed to reveal the intensity of cation-anion associations.\cite{wang_Superconcentrated_2016, yamada_Superconcentrated_2015, li_Stable_2021, bi_Cluster_2024, efaw_Localized_2023, jiang_Probing_2022, Chaffin_Solvation_2019, hahn_Influence_2020, jiang_Inhibiting_2021} NMR spectroscopy, a non-destructive and atom-specific technique, is particularly well-suited for investigating the molecular details of the chemical environments of the particular nucleus within solvation structures, since its signal is sensitive to the local chemical environment.\cite{deng_Natural_2015, hu_25Mg_2018, wan_Natural_2016, hu_Understanding_2022, jiang_Probing_2022, chen_Role_2020, im_Understanding_2022, allen_coordination_2024}. Apart from that, the experimental NMR spectroscopy reveals much more information about correlation times,\cite{hu_Situ_2018, allen_coordination_2024} relaxation times\cite{lin_Unravelling_2021,allen_quantifying_2023} and exchange dynamics of the chemical components.\cite{zhou_Probing_2015} Nevertheless, how to connect the observed spectral changes to their intrinsic molecular structural changes is a very challenging task. For instance, experimental NMR spectroscopy has observed chemical shift variations with changing concentration, particularly in the context of ion coordination in glyme electrolytes.\cite{yoshida_OxidativeStability_2011, plewa_glymeNMR_2010, wan_Natural_2016} While its solvation structural changes are hard to connect to the observed chemical shift. Some calculations, which employ density functional theory (DFT) calculation can provide some limited information into these trends for electrolytes, often employing cluster extraction with some complicated sampling methods.\cite{atwi_Automated_2022, chen_Role_2020, hu_25Mg_2018, deng_Natural_2015, im_Understanding_2022} There have been many static calculations attempting to explain the structure-spectrum relationship, using first-principles methods to calculate $^1$H, $^7$Li, $^{17}$O, $^{25}$Mg, $^{43}$Ca, $^{67}$Zn NMR spectra.\cite{deng_Natural_2015, wan_Natural_2016, hu_Understanding_2022, hu_25Mg_2018, hu_Modelling_2022, chen_Role_2020, atwi_Automated_2022, im_Understanding_2022} However, experimental chemical shifts reflect a weighted average from various local sites, combining local structural and dynamic information. This statistical averaging complicates signal resolution and makes it more challenging to evaluate the structure-spectrum relation.\cite{atwi_Automated_2022, lin_Machine_2022, lin_Unravelling_2021} 

Molecular dynamics (MD) simulations allow for tracking the dynamic structural changes in various types of electrolytes, with classical force field models,\cite{yao_Identifying_2024, atwi_Automated_2022, efaw_Localized_2023} first-principles methods\cite{goloviznina_Formation_2024, Perez_2020_Localized, yao_Applying_2022} and machine learning potentials.\cite{wang_Switching_2023, wang_Accelerating_2024} However, the direct connections between the molecular structures to the experimental spectral observables are still challenging due to the high computational cost of obtaining a single spectral response from the obtained MD configuration. Moreover, the delicate choices of configurational sampling when utilizing the NMR DFT calculation are too complicated to implement in the complex electrolyte system.\cite{atwi_Automated_2022} Nevertheless, there are some attempts, by training on structural descriptors of solid-state structures and their corresponding NMR chemical shifts, machine learning models can enhance the speed of chemical shift predictions while maintaining high accuracy.\cite{paruzzo_chemical_2018, cordova_solidMLNMR_2022, lin_Machine_2022, lin_combining_2022, xu_NMRnet_2024} Therefore, a robust computational method is essential, as a clear consensus on the relationship between the dynamic structural characteristics of electrolytes and their experimental spectral observables has not yet been established. Furthermore, validating the simulated structures and linking them to experimental observables remains a challenging task, one that could serve as a benchmark for the simulation.

In this paper, we propose a novel approach to connect the microscopic molecular structures and the spectroscopic features of the electrolyte. First, we combine two machine learning (ML) models for calculating dynamic NMR shifts: one MLP model for accelerating configuration sampling and another Neural Network (NN) model for rapid chemical shift prediction. To elaborate, we use the MLP model to generate a range of concentration trajectories. Subsequently, we conduct extensive sampling of electrolytes through prolonged MLMD simulations, creating chemical shift datasets from infrequent configurations within these simulations using the DFT method. The NN model is then trained and validated with this DFT-generated data, where the input features are the structural descriptor, such as Local Many-body Tensor Representation (LMBTR),\cite{huo_Unified_2022, dscribe2} and the outputs are the chemical shifts. \comment{Once the NN model is trained, the predictions become highly efficient, as the local structures are directly converted into descriptors, enabling rapid chemical shift prediction with minimal computational cost.} Finally, we use the NN-NMR model to calculate the dynamic $^7$Li NMR spectra for various concentrations of LiFSI in DME solutions. We successfully reproduce the anomalous reverse of $^7$Li NMR chemical shift observed in our NMR experiment when electrolyte concentration approaches the concentrated limit. Furthermore, by introducing one local structural parameter, called local structural index (LSI),\cite{shiratani_Growth_1996, shiratani_Molecular_1998, biswajitsantra_Local_2015} we quantitatively establish the structure-spectrum relationship, which helps clarify the solvation structure assignments. \comment{Moreover, we utilize the deformation factor ($\phi$) of the Electron Localization Function (ELF) at the Li nucleus to elucidate the impact of local structural order around Li$^+$ ions on electron shielding and consequently on NMR chemical shifts.\cite{savin_ELF_1992, Poater_ELF_2005, yao_Identifying_2024}} To this end, we find two different types of local solvation structures competing with each other, as electrolyte concentration approaches the concentrated limit, leading to the anomalous reverse of $^7$Li NMR chemical shift in our experiment. As such, we establish a comprehensive molecular-level understanding of the delicate microscopic molecular structures observed by using NMR spectroscopy. With our approach, one can track the complicated changes of solvation structures in the electrolytes with different conditions, enabling precise control over the stability and solubility properties of electrolytes.

\section{Method}
\subsection{Theoretical NMR Calculation}

\begin{figure}[htb] 
	\centering 
	\includegraphics[width=1.0\textwidth]{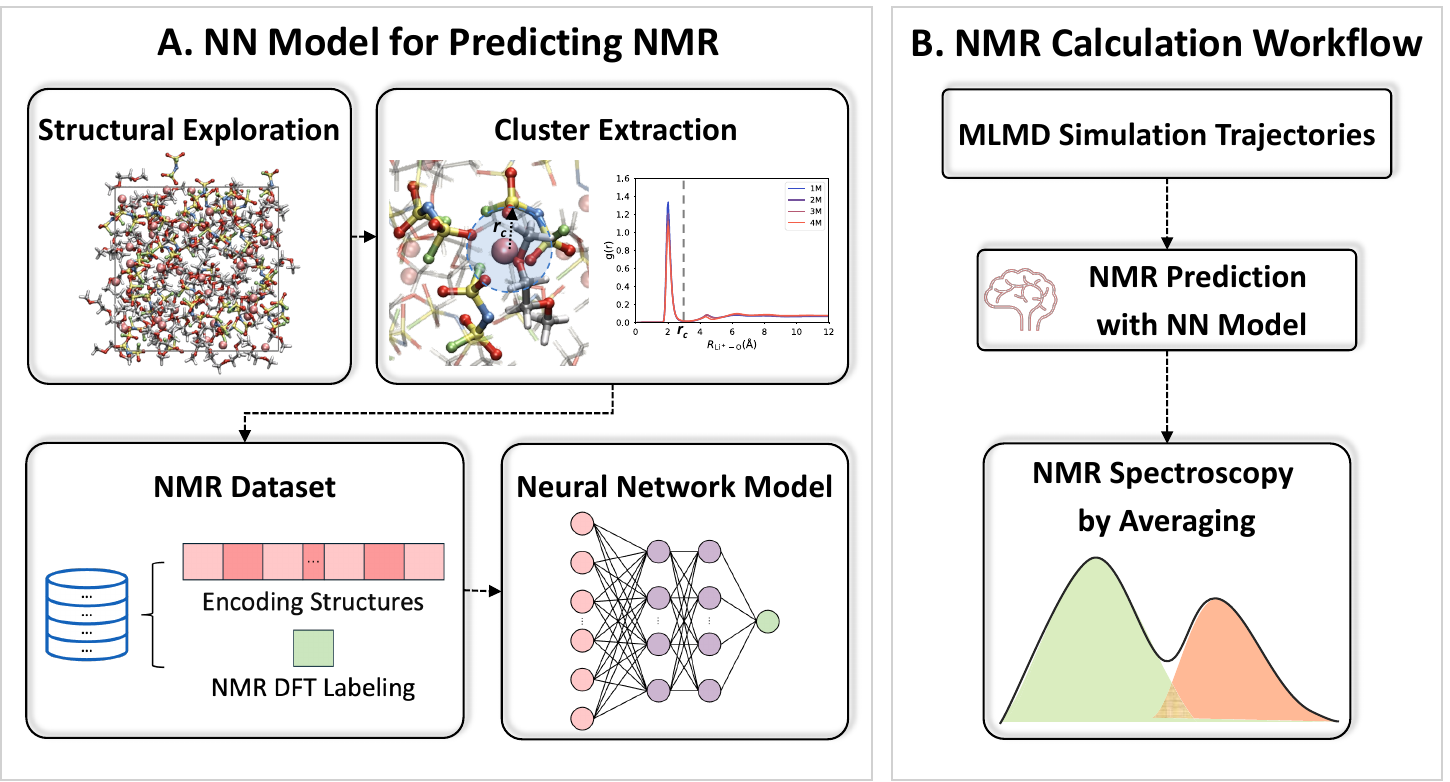} 
	\caption{Workflow for predicting NMR spectroscopy. (Left) Technique approach to training a neural network (NN) model. Explore various structures for different concentrations, and extract clusters surrounding lithium ions. Subsequently, encode structures using the descriptor and calculate their corresponding chemical shifts.
    (Right) NMR prediction workflow. Generate trajectories using MLMD simulations, then utilize the obtained NMR prediction NN model to obtain NMR spectroscopy.} 
	\label{fig:wf} 
\end{figure}

The methodology is illustrated in Fig.~\ref{fig:wf}. For the structural exploration section, Machine Learning Molecular Dynamics (MLMD) simulations are conducted under the NVT ensemble with various initial structures of different concentrations pre-equilibrated from classical MD simulations. The Nose-Hoover thermostat is employed to generate the NVT ensembles. The temperature is set to 300 K and each simulation lasts 10 nanoseconds (ns) with a 0.5 fs time step. More details about the pre-equilibrium and the MLMD simulations are given in Sect. I of the Supporting Information (SI). For the construction of the NN model, we sparsely sample configurations from MLMD trajectories. Then we extract the first solvation shells around the Li$^+$ ions as clusters and label them, resulting in a DFT dataset containing about 28,000 $^7$Li chemical shifts. All DFT calculations for NMR are performed by using Gaussian 16.\cite{g16} NMR calculations for clusters are conducted at the revTPSS\cite{tao_Climbing_2003, perdew_Workhorse_2009}/pcSseg-1\cite{jensen_Segmented_2015} level using the polarizable continuum model (PCM) solvation model.\cite{miertus_Electrostatic_1981}

The principle of cluster extraction is to capture the first solvation shell around the given Li$^+$ ion while maintaining the integrity of surrounding molecules. The first solvation shell is identified by the first minimum position $\sim$ 3.0 $\text{\AA}$ after the main peak in the radial distribution function (RDF) between Li and O shown in Fig. S3(c). We consequently extract complete molecules that have atoms located within this region. However, in some cases, other nearby Li$^+$ ions occupy the space within the first coordination shell, as shown in the RDF of Li$^+$-Li$^+$ at 4 M in Fig. S3(d), which exhibits a gentle peak around $\sim$ 3.0 $\text{\AA}$. Thus, we grasp the entire group of central atoms bonded to each other as one long-chain cluster. Although the cutoff distance between Li$^+$ ions is sufficiently set within the range of the first shell, we extend the distance to $\sim$ 6.3 $\text{\AA}$ to ensure the accurate local environment, which corresponds to the prominent peak position of RDF between Li$^+$ ions at 1-3 M.

For the construction of the NMR dataset, we use the LMBTR descriptor to encode the structure. The LMBTR vectors for the Li$^+$ local environments are generated using the Dscribe package\cite{dscribe2}, with a cutoff distance r$_{cut}$ of 6 $\text\AA$, the parameters of k2 and k3 are listed in Tab. S2 in SI. The NN prediction model is initialized and trained in PyTorch\cite{pytorch}, utilizing the hidden layer with three fully connected layers, each containing 256 nodes. The learning rate is initialized at 10$^{-3}$ with a total of 1000 epochs, using the Adam optimizer. The NMR dataset is divided into training, testing, and validation datasets in an 8:1:1 ratio, with an early stopping mechanism employed to prevent overfitting. We validate our NN-based NMR model using LiFSI/DME solutions. As shown in Fig. S5, the root-mean-square-error (RMSE) for the $^7$Li isotropic values in the testing dataset is $\sim 0.13$ ppm. Details of the NN model are provided in Sect. VI of SI. 

After obtaining the NN models, we predict the NMR spectra based on MD simulation trajectories. As shown in Fig.~\ref{fig:wf}(b), we generate four 10 ns MLMD trajectories for 1, 2, 3 and 4 M, respectively. We grasp the snapshots in regular intervals to ensure the number of Li$^+$ ions is approximately 90,000 for every concentration, which is enough to sample the chemical space. These snapshots are then encoded by the LMBTR descriptor as inputs for the NN model to predict NMR chemical shifts, which are aggregated into histograms of NMR chemical shifts corresponding to the respective concentrations. The Full Width at Half Maximum (FWHM) obtained from Lorentzian fitting for experimental data are 0.008, 0.017, 0.020 and 0.019 for 1 M, 2 M, 3 M and 4 M, respectively. We use these values to fit the associated histograms into NMR spectra, ensuring that the mean values of the histograms and the peak position of NMR spectra align (Fig.~\ref{fig:cs}(b)). The Lorentz function involved is: $Y(\omega) = A \frac{\gamma}{(\omega - \omega_0)^2 + \gamma^2}$,
where $\gamma$ is the FWHM, $\omega_0$ is the peak position, and A is a scaling factor.

\subsection{Experimental NMR Measurement}

Lithium bis(fluorosulfonyl)imide (LiFSI) and 1,2-dimethoxyethane (DME) are obtained from Duoduo Chemical Reagent Co., Ltd., with both chemicals having a purity of 99.99\%. All sample preparation procedures for four electrolytes with different salt concentrations (1, 2, 3, and 4 M LiFSI) are conducted within an argon-filled glovebox to maintain anhydrous conditions. $^7$Li Nuclear Magnetic Resonance (NMR) spectroscopy is employed to characterize LiFSI/DME electrolytes of varying concentrations. Measurements were performed using a Bruker Avance III HD 400 MHz NMR spectrometer at room temperature. We conduct three repeated NMR measurements for every single concentration. The experimental NMR chemical shifts data are shown in Fig.~\ref{fig:cs}(a). 

\section{Results and Discussion}
\subsection{NMR Spectra of LiFSI/DME Electrolytes with Different Concentrations}

\begin{figure}[htb] 
	\centering 
	\includegraphics[width=1.0\textwidth]{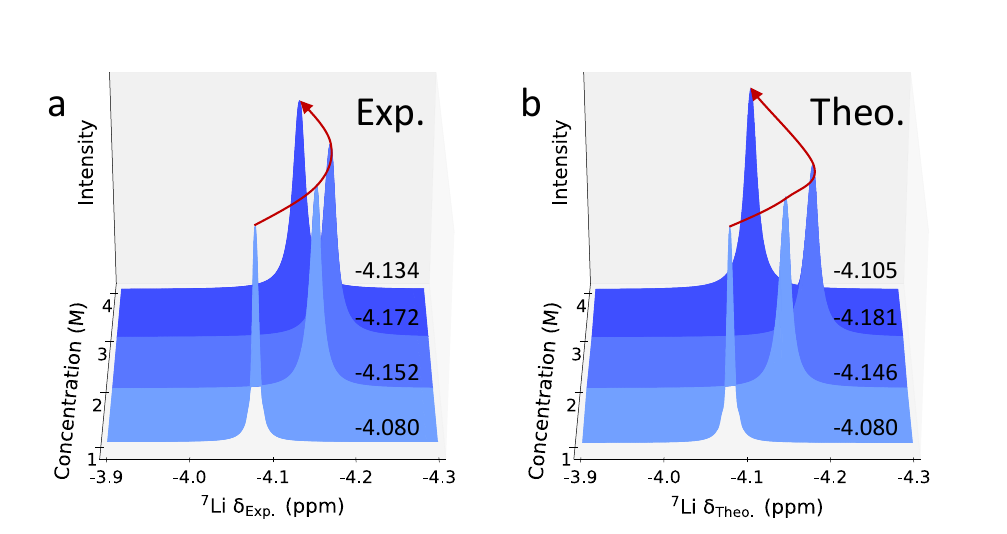}
	\caption{Comparison between computational and experimental results. (a) NMR experimental spectra of 1-4 M LiFSI/DME solutions. (b) Computed NMR spectra of 1-4 M LiFSI/DME solutions. The calculated average chemical shift of the lowest concentration electrolytes has been aligned with the experimental value. The 95$\%$ confidence interval values for experimental and theoretical NMR chemical shift are $\sim$ 0.006 ppm and $\sim$ 0.002 ppm, respectively. The red curves are provided as visual guides for showing the trend with increasing concentration.} 
	\label{fig:cs} 
\end{figure}

It is generally believed that in the dilute limit case, the solvated Li$^+$/FSI$^-$ ions are fully separated by the solution DME molecules in the LiFSI/DME electrolytes, called solvent-separated ions pairs (SSIP) situation. As the concentration of LiFSI increases, the coordination numbers of the oxygen atoms of DME molecules around the Li$^+$ ions change. More complicated solvation structures, like contact ion pairs (CIPs), and ion aggregates (AGGs) are emerged.\cite{xiao_Insights_2023} This, in turn, affects the ELF of the Li$^+$ ions,\cite{savin_ELF_1992, Poater_ELF_2005, krivdin_liquid_2023, yao_Identifying_2024} resulting in altered shielding effects around the Li$^+$ ions and changes in the observed NMR chemical shifts. This mechanism underlies the variations seen in the experimental NMR spectra for LiFSI/DME at different concentrations, as shown in Fig.~\ref{fig:cs}(a), our measurement is consistent with the previous reports\cite{wan_Natural_2016, jiang_Probing_2022}. As one can see, at the low concentrations (1-3 M), the NMR chemical shift tends to shift upfield with the increasing LiFSI concentration. However, at 4 M LiFSI concentration, which is close to the saturated limit ($\sim$5.3 M) of LiFSI/DME solution, the peak is downfield shifted to $-4.134$ ppm, when compared with the results obtained on the 3 M case of $-4.172$ ppm. The sudden changes of the chemical shift obviously reflect the changes of the Li$^+$ ion solvation structures at various LiFSI concentrations. 

To reveal the microscopic structural picture concerning the changes of the NMR chemical shift, we calculate the dynamic NMR spectra with the help of NN models, and the data is presented in Fig.~\ref{fig:cs}(b). Similarly to the experimental data, the theoretical NMR chemical shift moves upfield at low concentration, whereas it shifts downfield at 4 M. Notably, our theoretical NMR spectra are in excellent agreement with the experimental data, as we clearly show the NMR chemical shift turning point at 3 M concentration. The similar turning points observed in both experimental and theoretical NMR spectra suggest structural changes in the electrolytes as the LiFSI/DME solution approaches its saturation limit. Notably, at this concentration limit, the number of FSI$^-$ ions becomes comparable to that of the solvent molecules (DME), resembling a scenario referred to as `water-in-salts' (WiSs)~\cite{goloviznina_Formation_2024, zheng_Understanding_2018, wang_Switching_2023}. In this high-concentration regime, Li$^+$ ions are presumed to coordinate more frequently with FSI$^-$ ions than with DME molecules, in contrast to the low-concentration case. Since Li$^+$ ions tend to bond with oxygen atoms in either FSI$^-$ ions or DME molecules, the differing interaction strengths of the Li–O bonds with FSI$^-$ or DME disrupt the delicate balance of electron density around the Li$^+$ ions. This, in turn, affects the dynamic shielding effects and the NMR chemical shift. Nonetheless, the accurate reproduction of chemical shift variations in NMR due to changes in LiFSI concentration demonstrates the reliability of our NN models as effective tools for interpreting $^7$Li NMR experiments involving electrolytes.

\subsection{Local Structures of the LiFSI/DME Electrolytes}

\begin{figure}[htbp] 
	\centering 
	\includegraphics[width=1.0\textwidth]{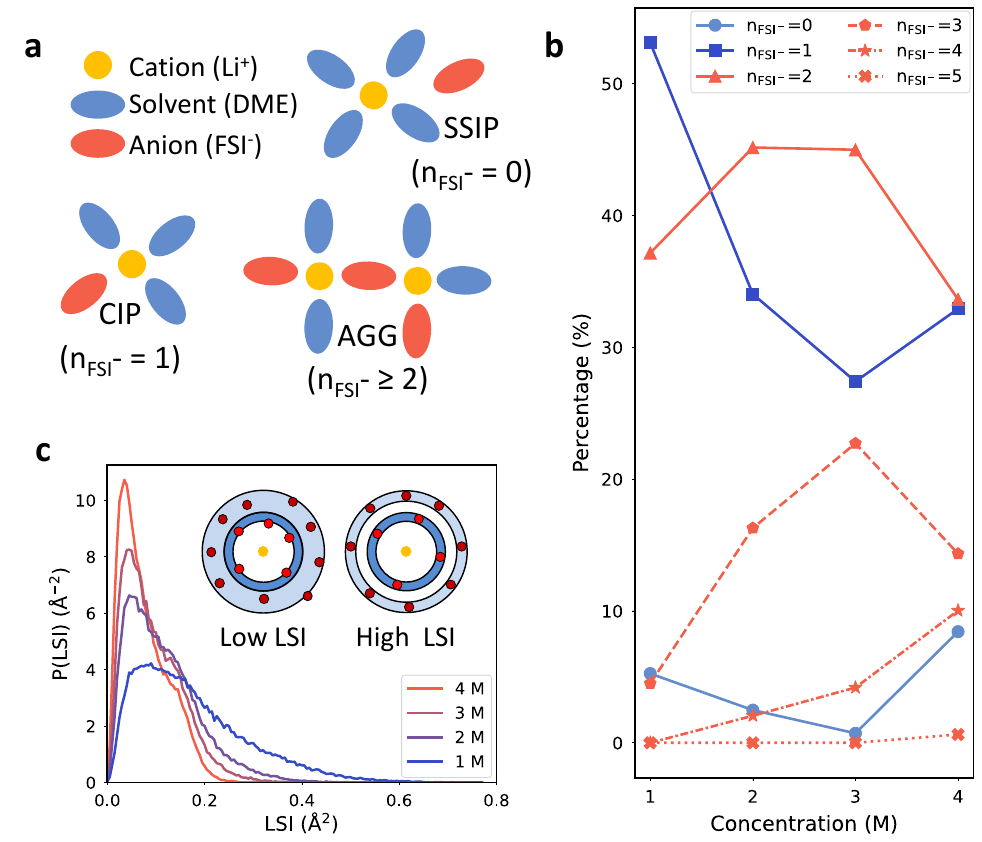} 
	\caption{(a) Schematic diagram of solvation structures. (b) Solvation structure categories for 1-4 M LiFSI/DME solutions. n$_{\rm FSI^-}$ means the number of FSI anions coordinate to the Li$^+$. (c) Probability density distributions of local structural index (LSI) for 1-4 M LiFSI/DME solutions. (Insert) Schematic description of high-density-like and locally disordered ({\it left}) {\it vs.} low-density-like and locally ordered ({\it right}) environments, which correspond to {\it low} and {\it high} values of the LSI order parameter, respectively. The dark and light blue areas correspond to the first and second solvation shells around the central Li$^+$ ions. The gaps between the coordinated oxygen atoms (red circles) in the two solvation shells are ambiguous for low LSI values and distinct for high LSI values.} 
	\label{fig:fsi-cate} 
\end{figure}

When increasing the concentration, the changes observed in the NMR spectra are a direct consequence of the evolving solvation structures. At the heart of this transformation lies the intricate interplay between cation-solvent and cation-anion interactions, which gives rise to distinct categories of the combinations of the cations, solvents and anions, such as SSIPs, CIPs, and AGGs, see Fig.~\ref{fig:fsi-cate}(a). Before interpreting the complicated chemical shift variations due to the changes in LiFSI concentration, we first focus on the classifications of the solvation structures, based on the number of anions coordinated with Li$^+$, as shown in Fig.~\ref{fig:fsi-cate}(b), which demonstrates that the high-number FSI$^-$ category (more than two) dominates as the concentration increases. Not surprisingly, in the low concentration case (1 M), the solvated Li$^+$/FSI$^-$ ions are separated by the solution DME molecules in the LiFSI/DME electrolytes, therefore, the coordination number of FSI$^-$ anions to Li$^+$ ion $n_{\rm FSI^-}$ is 0 or 1, which are corresponding to SSIP and CIP scenario. When the LiFSI concentration increases (2 or 3 M), the ratios of the SSIPs ($n_{\rm FSI^-}=0$) and CIPs ($n_{\rm FSI^-}=1$) decrease, and the Li$^+$ ions are more bonded with FSI$^-$ anions leading to the increase of the AGGs ($n_{\rm FSI^-}=2$ or 3). As more LiFSI salt is dissolved in the DME solution (4 M), approaching the saturation limit ($\sim$5.3 M)\cite{wan_Natural_2016}, the situation changes dramatically. The ratio of SSIPs and CIPs no longer continue to decrease but instead shows a slight increase compared to the 3 M case. Meanwhile, the proportion of lower-order AGGs ($n_{\rm FSI^-} = 2$ or 3, noted as AGGs) surprisingly decreases, while the proportion of higher-order AGGs ($n_{\rm FSI^-} > 3$, noted as AGGs+) continues to increase.

The terminology of SSIPs, CIPs, and AGGs can provide a framework for quantifying the solvation structures surrounding Li$^+$ ions. Meanwhile, the degree of inhomogeneity in the local molecular environment of Li$^+$ ions plays a crucial role in determining electron density around the Li$^+$ nucleus, which in turn influences dynamic shielding effects and the resulting NMR chemical shift. Here, we utilized an order parameter, which associates local structure index (LSI)\cite{shiratani_Growth_1996,shiratani_Molecular_1998,biswajitsantra_Local_2015} to the individual Li$^+$ ions and their local neighboring oxygen pairs. In short, the LSI order parameter is the mean-squared-deviation among the radial distances corresponding to the set of the oxygen atoms that surround a given Li$^+$ ion, and the LSI value is assigned as the inhomogeneity in the distribution of radial distances. The detailed definition can be referred to Sect. VII in SI. The schematic diagram provided in the insert of Fig.~\ref{fig:fsi-cate}(c) illustrates the different local atomic environments that can be distinguished and quantified by the LSI order parameter. For instance, a Li$^+$ ion with a high LSI value is typically found in a more ordered local environment, where neighboring oxygen atoms are densely concentrated around $\sim$ 2.0 $\text{\AA}$ and sparsely distributed around $\sim$ 3.0 $\text{\AA}$. This results in a clearer separation between the first and second coordination shells and a relatively low local atomic number density (as depicted on the right of the inset). Consequently, the electron density around the central Li$^+$ ion is less perturbed by the surrounding oxygen atoms, potentially leading to lower (absolute) NMR chemical shift values. Conversely, a Li$^+$ ion with a low LSI value resides in a locally disordered environment. This is characterized by a relatively high packing of neighboring oxygen atoms in the interstitial region and an elevated local atomic number density (as illustrated on the left of the inset). In such a densely packed and disordered atomic environment, the delicate balance of the surrounding electron density is disrupted, resulting in a higher NMR chemical shift value.

Now, let us focus on the impact of LiFSI concentration changes on the LSI distribution, as shown in Fig.~\ref{fig:fsi-cate}(c). At low concentrations, the LSI distribution exhibits a broad profile, suggesting a relatively ordered packing of neighboring oxygen atoms in the interstitial region. As the concentration of LiFSI increases in the DME solution, the interstitial region around the central Li$^+$ ion becomes more compressed. Consequently, the LSI distribution sharpens and displays a more pronounced peak. The distance distribution of the peak in 4 M spans from 1.8 $\text{\AA}$ to 3.0 $\text{\AA}$ with a bin size of 0.1 $\text{\AA}$. However, the distance distribution of the shoulder, which emerges at LSI =$\sim$ 0.15 $\text{\AA}^2$, is concentrated between 1.8 $\text{\AA}$ and 2.2 $\text{\AA}$, then vanished before reappearing beyond 3.0 $\text{\AA}$. The comparison means that the smaller the LSI value is, the tighter the first solvation shell is. As many more oxygen atoms are aggregating around Li$^+$ ions, distinct sub-interstitial structures occur between the first and the second coordination shell, which can be corroborated by the forward shift of the second coordination shell at 4 M, as shown in Fig. S3(c). This structure can also be analyzed using the RDFs between the Li$^+$ ions, see Fig. S3(d) of the SI. Notably, in the 4 M case, an additional peak appears around 3.0 $\text{\AA}$ in the Li$^+$-Li$^+$ RDF, indicating a direct interaction between two Li$^+$ ions. This suggests that the solvation structures of neighboring Li$^+$ ions are merging. Consequently, locally high-density AGGs+-like structures, also referred to as micelle-like structures\cite{efaw_Localized_2023, verma_Micellelike_2024}, begin to emerge, as illustrated in Fig.~\ref{fig:lsi-cs}(b) and (c). Under these conditions, the remaining Li$^+$ ions have a higher probability of forming ion pairs, such as SSIPs and CIPs. As shown in Fig.~\ref{fig:fsi-cate}(b), the ratio of SSIPs and CIPs no longer decreases but instead exhibits a slight increase compared to the 3 M case.

\subsection{Connections between Local Structures and NMR Chemical Shift}

\begin{figure}[htb] 
	\centering 
	\includegraphics[width=1.0\textwidth]{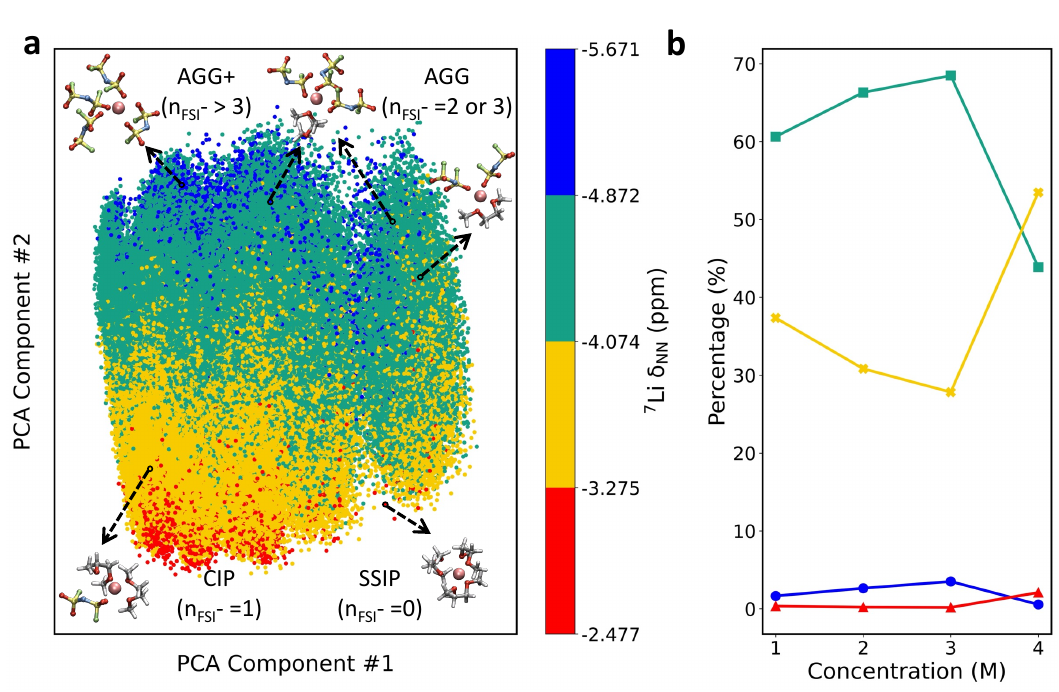}
	\caption{(a) Principal Component Analysis (PCA) of Li$^+$ solvation structural pattern mapping across various concentrations. Each point is associated with an NMR value from a local atomic motif, and the color code represents the corresponding NMR chemical shift value. The {\it x}- and {\it y}-axes represent the two most important principal components (PC\#1 and PC\#2). The representative fragments are colored as follows: red for SSIP, yellow for CIP, green for AGGs, and blue for AGGs with $n_{\rm FSI^-} = 3$ as well as AGGs+ ($n_{\rm FSI^-} > 3$). (b) The changes in the ratio of representative species for different colors for varying LiFSI concentrations.} 
	\label{fig:pca} 
\end{figure}

Our investigation focuses on elucidating the relationship between local structure and chemical shifts, emphasizing the need to unravel the complexities of the local environment and its corresponding isotropic characteristics. In the above analysis, we have interpreted the local chemical environment of Li$^+$ ions using chemical intuition, highlighting how the surrounding electron density is modulated by variations in the local environment. However, establishing a direct quantitative correlation between the semi-empirical relationship of the local chemical environment and the experimental NMR chemical shifts remains a more challenging task. To address this, we employ dimensionality reduction techniques, such as unsupervised principal component analysis (PCA). Specifically, we encode the local environment of the Li$^+$ ion using the LMBTR,\cite{huo_Unified_2022} a structural descriptor that captures the local structure and its corresponding spectral information in a high-dimensional physical latent space. We then apply PCA to reduce the dimensionality of this representation, as illustrated in Fig.~\ref{fig:pca}(a). 

\comment{Notably, contrary to the common observation that PCA often yields components with limited physical interpretability, we find that PC$\#$1 reflects intrinsic homogeneous structural information—such as solvation symmetry and the orientational changes of molecules surrounding Li$^+$—while PC$\#$2 captures the local chemical environment around Li$^+$ ions.} Specifically, points with the lowest NMR values (red) are typically associated with SSIPs ($n_{\rm FSI^-}=0$), while those with intermediate NMR values (yellow) correspond to CIPs ($n_{\rm FSI^-}=1$). The dominant chemical shift region (green) is linked to AGGs ($n_{\rm FSI^-}=2$ or 3), and the highest NMR values are observed for AGGs ($n_{\rm FSI^-}=3$) and AGGs+ ($n_{\rm FSI^-}>3$). The detailed assignment can be found in Tab. S4 of the SI. This finding aligns with chemical intuition, as discussed in the previous section, and underscores how the unsupervised PCA method effectively captures the influence of variations in the local environment on surrounding electron density.

Now, we can further investigate the changes in their populations to the varying LiFSI concentrations, as shown in Fig.~\ref{fig:pca}(b). Notably, a dramatic shift occurs between 3 M and 4 M. Clearly, the most dominant structures are those with intermediate NMR values (yellow) and those with higher NMR values (green). The former shows a sudden increase, while the latter experiences a sharp decrease. Building on the insights from the discussion from the previous section, we infer that the rise in CIP structures (yellow) is due to the emergence of highly localized AGGs+ structures induced by Li$^+$-Li$^+$ interactions. Moreover, the detailed decomposition of the PCA analysis for the different LiFSI concentrations, as shown in Fig. S7 of the SI, reveals similar distributions across the various concentrations. It clearly demonstrates the increased population of the CIPs as the concentration increases from 3 M to 4 M.

\begin{figure}[htb] 
	\centering 
	\includegraphics[width=1\textwidth]{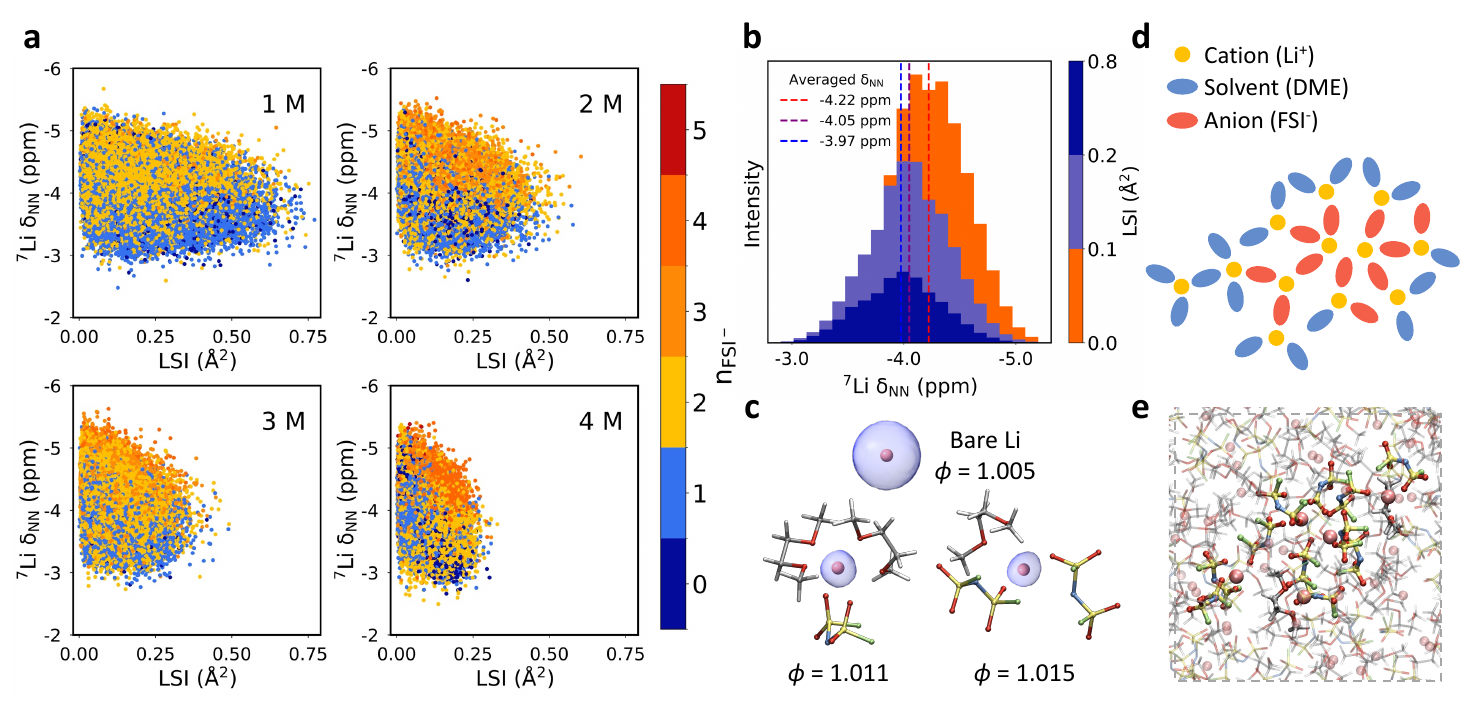} 
	\caption{\comment{(a) The correlation maps between the LSI and the chemical shift values for each concentration. The color bar represents the number of FSI$^-$ anions coordinated with the given Li$^+$ ions, denoted as n$_{\rm FSI^-}$, which determines the color of the points. (b) The histograms of chemical shifts within different LSI ranges, with the dashed red, purple, and blue lines representing the mean chemical shift corresponding to the 0$\sim$0.1, 0.1$\sim$0.2, and 0.2$\sim$0.8 $\text{\AA}^2$ LSI ranges. (c) Visualization of the electron localization function (blue) with different deformation factor $\phi$ around the Li nucleus (pink sphere). (d) Schematic illustration and (e) Snapshot of Li$^+$–FSI$^-$ long-chain cluster. The elements are colored as follows: Li in pink, C in gray, H in white, O in red, S in yellow, N in blue, and F in green.}}
	\label{fig:lsi-cs}
\end{figure}

Going further, the results from the PCA are supported by the local structural parameters, such as the n$_{\rm FSI^-}$ and LSI. In Fig.~\ref{fig:lsi-cs}(a), we present the NMR chemical shift values as a function of various LSI values, with the color assignment according to n$_{\rm FSI^-}$. The given Li$^+$ ion in a higher n$_{\rm FSI^-}$ solvation environment tends to move towards the upfield direction in chemical shift, and vice versa, which provides evidence for the region assignment of solvation structures in the PCA. \comment{Also, we split the LSI into three ranges and calculate the corresponding average chemical shifts, the histogram of which can be seen in Fig.~\ref{fig:lsi-cs}(b), illustrating that a smaller LSI range predominantly exhibits a lower chemical shift range.} As the chemical shift values become higher, the LSI values simultaneously reduce, the tendency of which is consistent in four concentrations. The reason for this phenomenon is that the higher-order coordination number ($n_{\rm FSI^-}\geq4$) configurations tend to have a filled sub-interstitial and form several long-chain clusters of Li$^+$ and FSI$^-$ connected, the LSI of which are lower than those of clusters with $n_{\rm FSI^-}<4$. Moreover, the clusters with $n_{\rm FSI^-}=0 $ or 1 (blue region) emerge along with the $n_{\rm FSI^-}=4$ case at 4 M, as shown in Fig.~\ref{fig:lsi-cs}(d) and (e), surrounding these AGGs+-like structures, ultimately leading to the lower LSI compared with 3 M case. \comment{Additionally, we investigate several solvation structures ranging from 1 M to 4 M to quantify the deformation factor $\phi$ of ELF of the Li nucleus,\cite{Multiwfn2012, Multiwfn2024, ChimeraX} the average of which for each concentration is about 1.0130±0.0002, 1.0128±0.0002, 1.0127±0.0001, and 1.0130±0.0001, respectively. The detailed definition and specifics can be found in Sect. IX of the SI. A schematic depicting the varying degrees of deformation is presented in Fig.~\ref{fig:lsi-cs}(c). As shown, the ELF deformation becomes more pronounced as the deformation factor increases. When deformation is more significant, shielding around the given Li nucleus weakens, leading to a downfield shift in the chemical shift.\cite{yao_Identifying_2024} Therefore, the average deformation factor experiences a decrease from 1 M to 3 M and an increase from 3 M to 4 M, which is consistent with the upfield shift of NMR chemical shifts, followed by the downfield shift.} 

\begin{figure}[htb] 
	\centering 
	\includegraphics[width=1\textwidth]{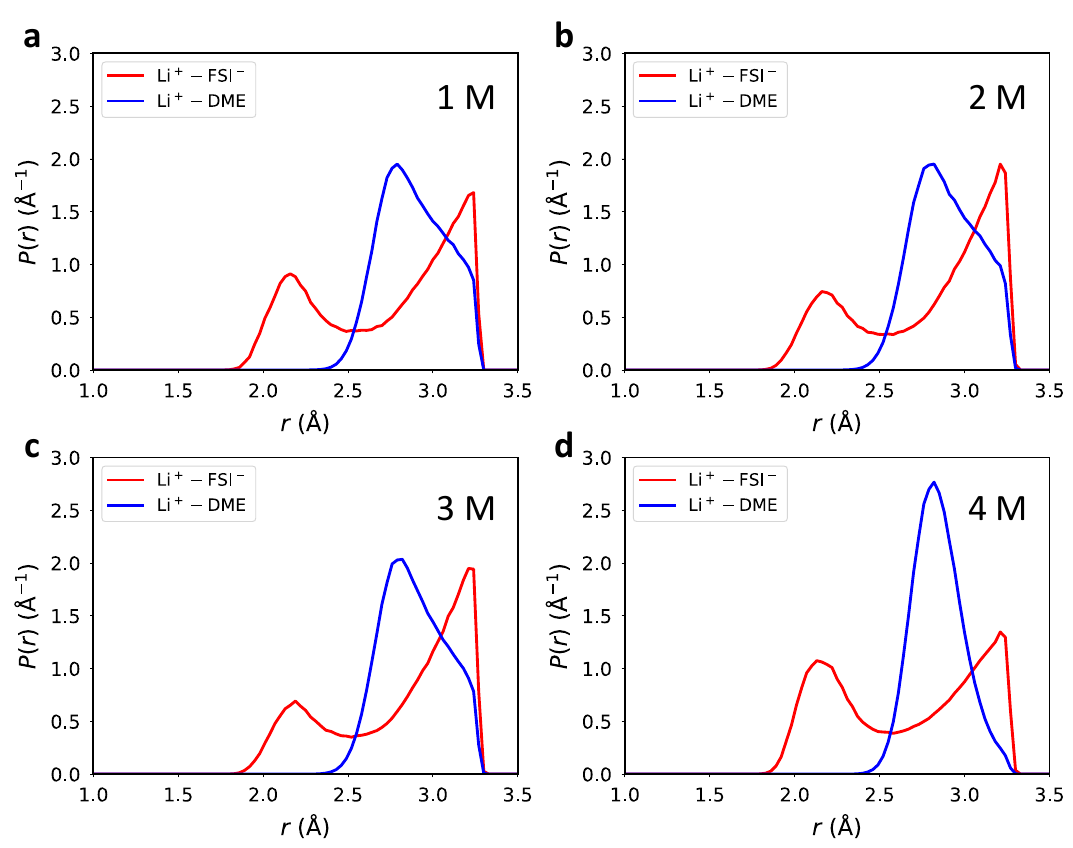} 
	\caption{Probability density of the molecule appearance at a particular distance within a 3.3 $\text{\AA}$ radius from the central Li$^+$ ions for (a) 1 M, (b) 2 M, (c) 3 M, (d) 4 M. The red line is for Li$^+$-FSI$^-$ pair and the blue line is for  Li$^+$-DME pair.} 
	\label{fig:separated-phase} 
\end{figure}

The long-chain clusters shown in Fig.~\ref{fig:lsi-cs}(d) and (e) tend to introduce the phase separation of DME and FSI$^-$. We have analyzed the surrounding molecules around the central Li$^+$ ions within 3.3 $\text{\AA}$, using the midpoint of each molecule for measurement. The curves illustrated in Fig.~\ref{fig:separated-phase} represent the probability density of the molecule appearance at a particular distance from the central Li$^+$ ions. They are similar in 1 to 3 M for both DME and FSI$^-$, where the main peak of FSI$^-$ appears at 3.15 $\text{\AA}$, and DME shows a shoulder around 3.15 $\text{\AA}$. In 4 M solutions, however, more FSI$^-$ anions penetrate the first solvation shell at $\sim$ 2.15 $\text{\AA}$, and DME molecules aggregate $\sim$ 2.80 $\text{\AA}$ with the shoulder disappearing. The aggregation of FSI$^-$ ions in the inner shell and the clustering of DME molecules create distinct phases in solutions, resulting in localized molecular redistribution and aggregation. The corresponding schematic diagrams are shown in Fig. S8 of the SI. This trend aligns with the findings from the local structure analysis in the previous section, which highlights the complex interplay between different local environments, with notable shifts driven by concentration effects. Simultaneously, the ELF effects arise due to the strong electrostatic interactions and localized electron density around Li$^+$ ions, which restricts the free movement of charge carriers. Meanwhile, the steric effects hinder the spatial pathways for ion migrations. These combined effects lead to a high viscosity and low conductivity in HCEs.\cite{peng_conduct_2018} Nonetheless, the distances between Li$^+$ ions are much closer in 4 M than the other three concentrations, reducing the distance required for the transportation of Li$^+$ ions to maintain the flux when Li$^+$ ions near the electrode are depleted at high current density. Also, the solid electrolyte interphase (SEI) formed in 4 M LiFSI/DME shows a slow anion degradation during extended cycling, resulting in a substantial accumulation of inorganic components.\cite{qian_High_2015} These components enhance ionic conductivity and provide mechanical protection, strengthening the SEI and significantly improving the CE when using a Li metal anode.\cite{qian_High_2015, wan_Multinuclear_2017} Nevertheless, the downfield shift of the $^7$Li NMR chemical shift from 3 M to 4 M, is the signature of the emergence of the highly localized AGGs+ structures and results in less coordinated CIPs structures in the high-concentration electrolytes.

\section{Conclusion}
To summarize, this study presents a novel machine learning-based approach for calculating dynamic NMR shifts in LiFSI/DME solutions by integrating an MLP model for configuration sampling and an NN model for chemical shift prediction. The NN-NMR model demonstrated both high accuracy and efficiency, with predictions closely matching experimental NMR spectra. \comment{This method can be applied to other complex electrolyte systems and extended to predict shifts for other nuclei. However, its accuracy depends on the quality of the training data, and more complex systems may require further adjustments, with future work focusing on expanding the dataset and refining the model.} As the concentration of LiFSI increases from 1 M to 3 M, changes in the solvation structure result in upfield shifts of the NMR chemical shifts, while at 4 M, the shifts move downfield. Through advanced modeling techniques, we build a quantitative relationship between molecular structure and NMR spectra, providing deep insights into solvation structure assignments. Our findings reveal the coexistence of two competing local solvation structures that exchange in dominance as electrolyte concentration approaches the upper limit, leading to observable changes in $^7$Li NMR chemical shifts. This approach provides valuable insights into the relationship between solvation structure and NMR shifts, offering a more efficient and insightful method for studying electrolyte solutions. Overall, this work enhances our understanding of electrolyte solvation and opens new pathways for optimizing electrolyte design based on molecular-level NMR insights.

\section{Associated Content} 
\subsection{Data Availability Statement}

The code of automated workflow (ai2-kit) can be found at https://github.com/chenggroup/ai2-kit, and datasets for NN can be found at \url{https://dataverse.ai4ec.ac.cn/dataset.xhtml?persistentId=doi:10.12463/AI4EC/3F9AOZ}.

\subsection{Supporting Information}
The Supporting Information is available free of charge at https://pubs.acs.org/xxx. Details on training errors of NN, AIMD setup, comparison between AIMD and MLMD, as well as NMR computational setup.

\subsection{Author Contributions}
\#Q. Y. and Y. S. contributed equally to this work.
\subsection{Notes}
The authors declare no competing financial interest.

\begin{acknowledgement}
 We thank Prof. Bing-wen Hu at East China Normal University and Prof. Hai-ming Liu at ShanghaiTech University for helpful discussions. F. T. acknowledges the National Key R\&D Program of China (Grant No. 2024YFA1210804) and a startup fund at Xiamen University. J. C. acknowledges the National Natural Science Foundation of China (Grant Nos. 22021001, 22225302, 21991151, 21991150, 92161113, and 20720220009) and the Laboratory of AI for Electrochemistry (AI4EC) and IKKEM (Grant Nos. RD2023100101 and RD2022070501) for financial support. This work used the computational resources in the IKKEM intelligent computing center.
	
\end{acknowledgement}

\bibliography{property}

\end{document}